\documentclass[manuscript,screen,review=false]{acmart}
\usepackage{multirow}

\AtBeginDocument{%
  \providecommand\BibTeX{{%
    \normalfont B\kern-0.5em{\scshape i\kern-0.25em b}\kern-0.8em\TeX}}}

\setcopyright{acmlicensed}
\copyrightyear{2024}
\acmYear{2024}
\acmDOI{XXXXXXX.XXXXXXX}

\acmConference[Conference acronym 'XX]{Make sure to enter the correct
  conference title from your rights confirmation emai}{June 03--05,
  2018}{Woodstock, NY}
\acmISBN{978-1-4503-XXXX-X/18/06}

\begin{document}

\title[SecureRights]{SecureRights: A Blockchain-Powered Trusted DRM Framework for Robust Protection and Asserting Digital Rights}

\author{Tiroshan Madushanka}
\authornote{Both authors contributed equally to this research.}
\email{tiroshanm@kln.ac.lk}
\orcid{0000-0001-6627-1477}
\author{Dhammika S. Kumara}
\authornotemark[1]
\email{dhammika.dsk@gmail.com}
\affiliation{%
  \institution{University of Kelaniya}
  \city{Kelaniya}
  \country{Sri Lanka}
}

\author{Atheesh A. Rathnaweera}
\affiliation{%
  \institution{University of Kelaniya}
  \city{Kelaniya}
  \country{Sri Lanka}
\email{rathnaweeraatheesh72@gmail.com}
}

\renewcommand{\shortauthors}{Tiroshan and Dhammika, et al.}

\begin{abstract}
  In the dynamic realm of digital content, safeguarding intellectual property rights poses critical challenges. This paper presents "SecureRights," an innovative Blockchain-based Trusted Digital Rights Management (DRM) framework. It strengthens the defence against unauthorized use and streamlines the claim of digital rights. Utilizing blockchain, digital watermarking, perceptual hashing, Quick Response (QR) codes, and the Interplanetary File System (IPFS), SecureRights securely stores watermark information on the blockchain with timestamp authentication. Incorporating perceptual hashing generates robust hash tokens based on image structure. The addition of QR codes enhances the watermarking, offering a comprehensive solution for resilient intellectual property rights protection. Rigorous evaluations affirm SecureRights' resilience against various attacks, establishing its efficacy in safeguarding digital content and simplifying rightful ownership assertion.
\end{abstract}


\keywords{ Digital rights management (DRM), Watermark, Blockchain, Copyright Protection, Ownership Assertion}

\received{20 February 2007}
\received[revised]{12 March 2009}
\received[accepted]{5 June 2009}

\maketitle

\section{Introduction}
\label{lbl:introduction}

In the rapidly evolving digital technology landscape, the exponential proliferation of multimedia and digital content on the internet has become a defining characteristic of the contemporary era. Digital images, artworks, soundtracks, and videos, once confined to traditional forms of dissemination, now inundate online platforms, facilitating unprecedented levels of sharing and accessibility for creators and consumers alike. However, this digital revolution has engendered a critical conundrum: the inherent vulnerability of digital content to unauthorized manipulation, reproduction, and distribution. The seamless nature of these processes poses a formidable risk to preserving intellectual property rights and information integrity, as malevolent actors exploit these vulnerabilities to infringe upon the legitimate rights of copyright holders and assert wrongful ownership.

In response to these challenges, legislative and regulatory bodies have introduced rules, patents, and licensing frameworks to safeguard digital content creators' ownership and intellectual property rights. Despite these efforts, the efficacy of such measures remains limited, necessitating innovative solutions to address the ever-expanding scope and intricacies of digital content protection. One such solution that has garnered attention is the concept of Digital Watermarking, initially proposed to protect the copyrights of electronic documents and subsequently extended to shield software copyrights by embedding watermarks within various data structures~\cite{Ibrahim2009}. However, digital watermarks are vulnerable to subtract attacks because the attacker can delete the embedded digital watermark content from the data structure without affecting the functionality. 

Within the realm of watermarking solutions, the semi-fragile watermarking scheme stands out as a prominent choice, involving the embedding of watermark(s) within the Discrete Wavelet Transform (DWT) domain. By embedding watermarks in the quantized DWT domain, this approach yields a robust scheme resilient to JPEG compression exceeding 50\% Quality Factor (QF) and achieves a minimum 32×32 detection unit~\cite{Kundur1999}. Moreover, it offers the possibility of inserting a signature extracted from the original image into the DWT domain. However, this signature proves less robust in the face of image processing and JPEG compression below 60\% QF~\cite{Zhou2004}. Discriminating between malicious and non-malicious attacks can be facilitated by integrating the Just Noticeable Differences (JND) feature into the embedding scheme~\cite{Kang2003}. Extracting image features from the low-frequency domain to generate two watermarks enables the classification of intentional content modifications and indicates modification locations. However, this approach is less robust to JPEG compression below 50\% QF~\cite{Hu2005}. Zernike moments in the DWT domain serve as robust features for authentication tasks, particularly effective against JPEG compression above 50\% QF and achieving a 32×32 detection unit~\cite{Liu2005}. Additionally, embedding inter-block and intra-block signatures in the DWT domain using a block-mean-based quantization scheme proves robust against JPEG compression exceeding 60\% QF~\cite{Zhu2007}. Despite the considerable volume of research devoted to safeguarding digital rights using digital watermarking techniques, a comprehensive framework for protecting and asserting digital copyrights remains notably absent. 

The proposed approach addresses this critical gap by proposing an innovative blockchain-based digital rights management system. By seamlessly integrating digital watermarking, blockchain technology, perceptual hashing, Quick Response (QR) codes, and the Interplanetary File System (IPFS), this system establishes a comprehensive platform for robust intellectual property (IP) rights protection. One of the primary challenges it tackles is the vulnerability of digital content to copyright violations, including unauthorized editing, misuse, and distribution. Through secure storage of watermark information on the blockchain and provision of timestamp authentication for multiple watermarks, the proposed system ensures a resilient defence against such violations. Moreover, the system enhances its watermarking capabilities by leveraging perceptual hashing to generate hash tokens based on structural image information and incorporating QR codes containing the image's hash token and copyright message. Additionally, managing digital content and copyright information within the IPFS Network further strengthens its holistic approach to IP rights protection. 

In summary, this study introduces an innovative blockchain-based digital rights management system, strategically combining digital watermarking, blockchain technology, perceptual hashing, QR codes, and IPFS. Addressing unauthorized content editing and distribution issues, the proposed framework demonstrates resilience against various common digital content attacks. The subsequent sections of this paper are organized as follows: Section~\ref{lbl:related_work} reviews existing research on Blockchain and Digital Watermarking in DRM, Tampering and Detection techniques. Section~\ref{lbl:proposed_method} details the novel blockchain-based digital rights management system, and Section~\ref{lbl:experiments} presents experimental findings on different digital content attacks, validating the framework's robust defence against copyright violations. Finally, Section~\ref{lbl:conclusion} offers concluding remarks.

\section{Related work}
\label{lbl:related_work}
In this section, we conduct an in-depth examination of blockchain applications in copyright management, digital watermarking techniques, and recent research findings. We delve into various aspects, including methods for detecting image tampering, integrating blockchain technology with Digital Rights Management (DRM), and using image watermarking for copyright protection. 

\subsection{Block-chain with Digital Rights Management}

Blockchain technology has been employed in various Digital Rights Management (DRM) applications. For instance, a Content Distributed System utilizes blockchain to manage access rights to 4K video content, enabling copyright owners to regulate permissions efficiently~\cite{Kishigami2015}. PeerTracks music platform leverages the Muse blockchain to facilitate streaming and retail music sales, incentivizing listener engagement through cryptocurrency rewards~\footnote{https://museblockchain.com/}. Similarly, Ujo Music~\footnote{https://ujomusic.com/} utilizes the Ethereum blockchain for music publishing and sharing, with functionalities including payments, rights management, and artist identity storage. At the same time, the Interplanetary File System (IPFS)~\footnote{https://ipfs.tech/} handles data transfer. Furthermore, blockchain technology has found application in software license validation, exemplified by the Master Bitcoin Model, where consumers demonstrate ownership by sharing bitcoins originating from the software vendor. The Bespoke Model extension allows software owners to embed additional license metadata, such as expiration dates, within bitcoins to enhance license management~\cite{Herbert2015}.

\subsection{Image Watermarking for Copyrights Protection}

In image watermarking for copyright protection, the imperative is to create an invisible watermark that remains robust against typical image processing techniques such as filter usage, contrast adjustments, and data transformations like JPEG conversion. Literature discerns two primary categories in image blind watermarking algorithms, namely Spatial Domain Techniques~\cite{Nikolaidis1998} and Frequency Domain Techniques~\cite{Cox1997}. Spatial domain techniques directly manipulate pixel values, often categorized as either differential expansion~\cite{Thodi2007} or employing histogram modifications~\cite{Coatrieux2013, Zong2015}. Frequency-domain techniques, on the other hand, modify the image's frequency components with the Discrete Cosine Transform (DCT)~\cite{Cox1997} and the Discrete Wavelet Transform (DWT)~\cite{Nasir2012} being prevalent. Spread-spectrum watermarking~\cite{Cox1997} treats the frequency domain as a communication channel, enhancing resistance to attacks but compromising visual quality. Techniques like embedding in high-frequency sub-bands~\cite{You2010} and placing watermarks in the lowest frequency components~\cite{Lee2014} aim to improve image quality and resistance to cropping attacks. Some methods employ a combination of singular value decomposition (SVD) and frequency transforms for a balance between robustness and invisibility~\cite{Lai2010}, albeit at the cost of increased computational complexity. A novel approach involving the division of DCT coefficients into vertical, horizontal, and diagonal directions~\cite{Lou2014} determines the dominant direction based on total energy, but reliability inconsistencies pose challenges. In image watermarking, navigating the trade-off between watermark robustness and invisibility is an inherent consideration.

\subsection{Tampering and Detection}
The landscape of digital image tamper detection encompasses active detection techniques (Non-Blind) and passive detection techniques (Blind), each with distinct characteristics and applications. Active approaches necessitate the presence of a watermark during the examination. They can be further categorized into (1) methods utilizing a fragile watermark, which detects tampering after image localization but faces challenges in distinguishing specific alterations, and (2) methods employing a semi-fragile watermark, capable of detecting significant changes while allowing content-preserving processing. A notable spatial domain, a block-based embedding technique for tamper detection and localization has been proposed by Bhalerao et al.~\cite{bhalerao2019block}.

Conversely, passive detection methods do not require pre-embedding watermarks or digital signatures and are widely used for tamper detection in digital images. These techniques include Copy-move forgery detection (Cloning) and Splicing. Mishra et al.~\cite{mishra2013review} employ an edge-blurring technique to detect various operations, while Prakash et al.~\cite{prakash2012image} propose a method using dyadic wavelet transform (DyWT). Splicing, the advanced merging of two or more images, is addressed by Iakovidou et al.\cite{iakovidou2021passive} introducing passive tampering localization methods for splicing on JPEG images. These passive techniques represent significant advancements in tamper detection, as they eliminate the need for prior information or pre-embedding, making them highly valuable in the continuous development of tamper detection methodologies. 

\section{Proposed Framework}
\label{lbl:proposed_method}
In this section, we present a novel framework to safeguard the copyrights of digital image content by integrating blockchain technology and watermarking techniques. We commence by delineating the problem statement elucidating the challenges faced in preserving digital image copyrights. Subsequently, we provide a comprehensive overview of our proposed Digital Rights Management (DRM) framework, outlining its essential components and functionalities. Finally, we delve into the architecture of the framework, elucidating its design principles and underlying mechanisms.

\subsection{Problem definition}
The ease of downloading digital works facilitates unrestricted distribution, enabling widespread piracy and unauthorized use that severely compromises the interests of content creators. This issue is particularly acute in photography, design, and e-commerce, where piracy hackers cause substantial financial losses. Content owners are compelled to invest significant time and resources in legal battles to establish protection against tampering and unauthorized alterations. Critical challenges within existing digital rights management systems compound the problem: (1) digital content is freely accessible without adequate download restrictions, (2) users struggle to authenticate content from often unknown sources, and (3) content providers find it challenging to track piracy effectively. The influence of multimedia content on social media further complicates matters as manipulated digital content proliferates, eroding the reliability of visual information and necessitating enhanced authentication mechanisms. In light of these pressing issues, this work addresses the urgent need for a robust framework to safeguard digital image copyrights in an environment rife with piracy, tampering, and the unrestrained distribution of digital content.

\subsection{SecureRights}
The proposed framework (i.e., SecureRights) is designed to enhance copyright protection for digital images, ensuring a secure environment for owner data. The framework employs blockchain innovations to fortify copyright data elements, delivering robust security. The process enhances accuracy while simplifying watermark execution and extraction by incorporating secure algorithms based on encryption and hash functions. The framework is divided into two main components.

The first part of the proposed CRM system encompasses hash generation, blockchain storage, digital watermark image generation, and digital watermark embedding. The perceptual hash function assigns a hash value as the image ID, ensuring consistency before and after watermark embedding. A cryptographic hash function is also employed to prevent confusion between the original and watermarked images. Blockchain technology securely stores image information, including the perceived hash value, image owner details, and image structural information. The decentralized nature of blockchain reduces the risk of tampering, and its timestamp function helps sequence multiple watermarks. Upon blockchain recording, the final stage involves generating and embedding a digital watermark using a QR Code that includes the owner's digital signature. The embedding process employs a frequency domain method, specifically the discrete cosine transform (DCT), to enhance watermark toughness and information capacity.

The second part addresses storage and dissemination, utilizing the Interplanetary File System (IPFS) to store and distribute watermarked images. The user initiates the process by uploading the original image and copyright owner information, generating digital signatures, creating a QR Code image as a watermark, and embedding it in the original image. This watermarked image and relevant information are uploaded to the IPFS network for browsing and downloading.

To verify the subjected image, the proposed framework utilizes the perceptual hash function to calculate the watermark image, extract the watermark, and compare hash values with blockchain records to determine copyright authenticity. This comprehensive methodology ensures the integrity of digital images, resolves issues related to multiple watermarks, and provides a secure and verifiable system for protecting digital image copyrights.

\subsection{Proposed Framework Architecture}

\begin{figure}
\centerline{\includegraphics[width=0.9 \textwidth]{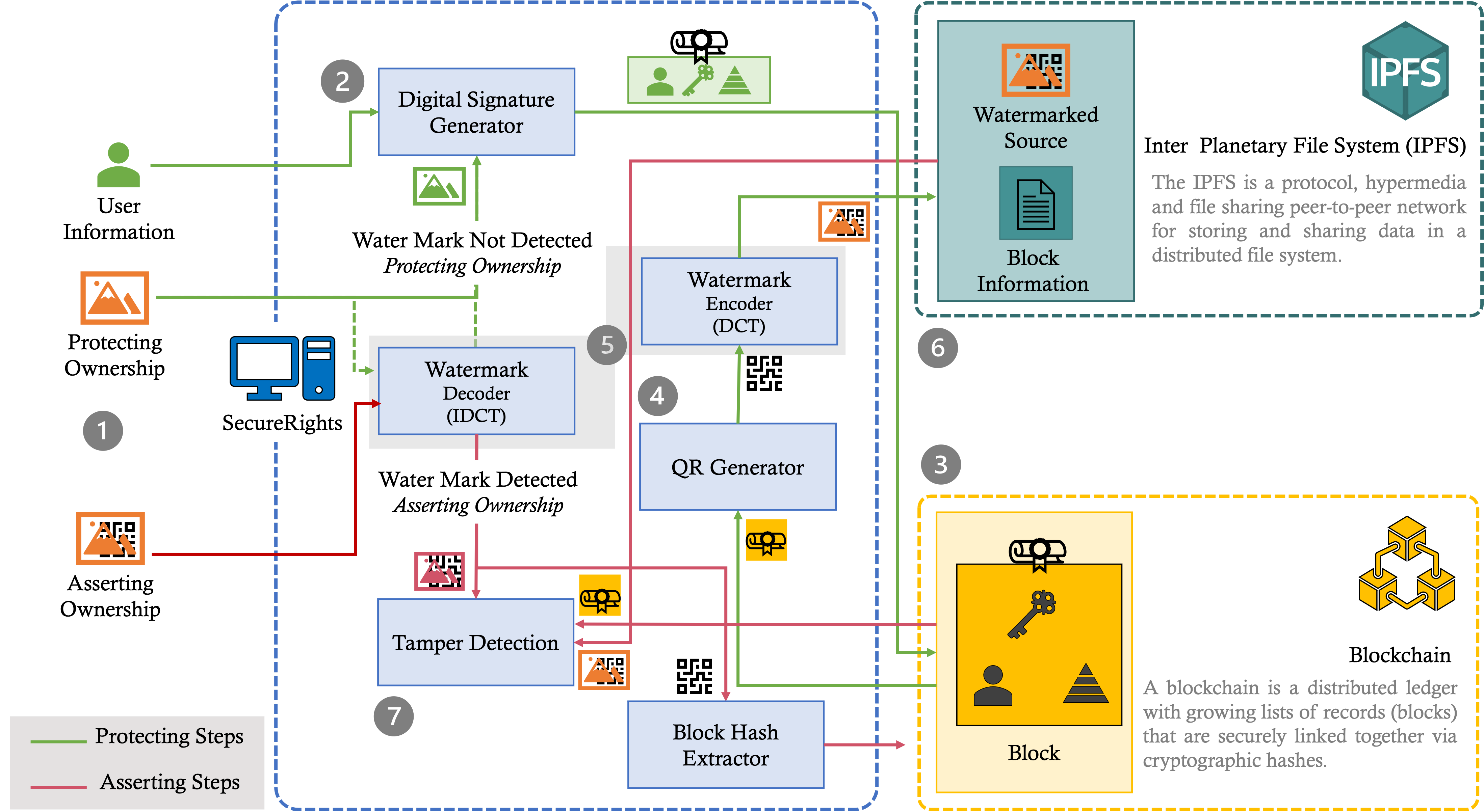}}
\caption{SecureRights: Proposed framework for digital content protection and asserting.}
\label{fig_architecture}
\end{figure}

The envisioned architecture, depicted in Figure~\ref{fig_architecture}, outlines the comprehensive framework designed to safeguard the copyrights of digital image content. The intricate functionalities of key components are elucidated below:

\subsubsection{Content Upload}
The primary step (i.e., step 1) in securing or asserting ownership requires the user to upload the intended digital image to the framework, providing essential data such as name, email, address, and picture metadata through the SecureRights Portal.

\subsubsection{Hashing and Digital Signature Generation}
The proposed framework utilizes the perceptual hash function in the hashing process to generate a unique image token number that remains unchanged with the digital watermark embedding. This hash token, a numeric value, cannot be traced back to the original message information. Perceptual hash functions are generally categorized into four types: Average Hash (AHA), Different Hash (DHA), Perceptual Hash (PHA), and Wavelet Hash (WHA). We have opted for the Different Hash (DHA) in our framework, which produces a 16-bit hexadecimal value.

In step 2, the perceptual hash value, information from the content owner, and structural information collectively serve as the digital signature for the copyright (content) owner.


\subsubsection{Block-chain Creation}
Referring to the generated digital signature, in step 3, the framework initiates the blockchain transaction and creates a data block in the blockchain, referencing the generated digital signature. Table~\ref{tab_blockchain} provides the details of the SecureRights DRM blockchain data instance for the Lenna image as illustrated in Figure~\ref{fig_watermark_embedding_and_extracting}(a).

\begin{table*}[htbp]
\caption{SecureRights DRM blockchain data instance}
\begin{center}
\begin{tabular}{ll}
\hline
Name              & Data                                                                                                                                                                                                                                                                                                                                                                       \\ \hline
Address           & \texttt{0x4598f10fa5757674a63d4f424913287961d3df44}                                                                                                                                                                                                                                                                                                                                 \\
Block number      & \texttt{605417}                                                                                                                                                                                                                                                                                                                                                                     \\
Block hash        & \texttt{0x8065f1948301a8ab627b58346629df699912db507eb71beec0677281e55bb2d1}                                                                                                                                                                                                                                                                                                         \\
Log index         & \texttt{7}                                                                                                                                                                                                                                                                                                                                                                          \\
Transaction hash  & \texttt{0xcc0f64fb86c307dd5c76262fe2f8defcd98524aacab876f94743fb0a4b5932ed}                                                                                                                                                                                                                                                                                                         \\
Transaction index & \texttt{6}                                                                                                                                                                                                                                                                                                                                                                          \\
Args              & \begin{tabular}[c]{@{}l@{}}\{\\    "Owner":"\texttt{0x51balbc194580887ce1d305774c58c678e25cb86}",\\    "ImageHash":"\texttt{a9e7c071d11a2aec6a8c8100061985d}",\\    "CreationName":"Lenna Image",\\    "CreationAuthor":"Tiroshan Madushanka",\\    "CopyrightOwner":"University of Kelaniya",\\    "ImageID":"\texttt{7670795b33135a38}",\\    "MailAddress":"tiroshanm@kln.ac.lk"\\ \}\end{tabular} \\ \hline
\end{tabular}
\label{tab_blockchain}
\end{center}
\end{table*}

\subsubsection{Quick Response (QR) Code Generation}

\begin{figure}
\centerline{\includegraphics[width=7cm]{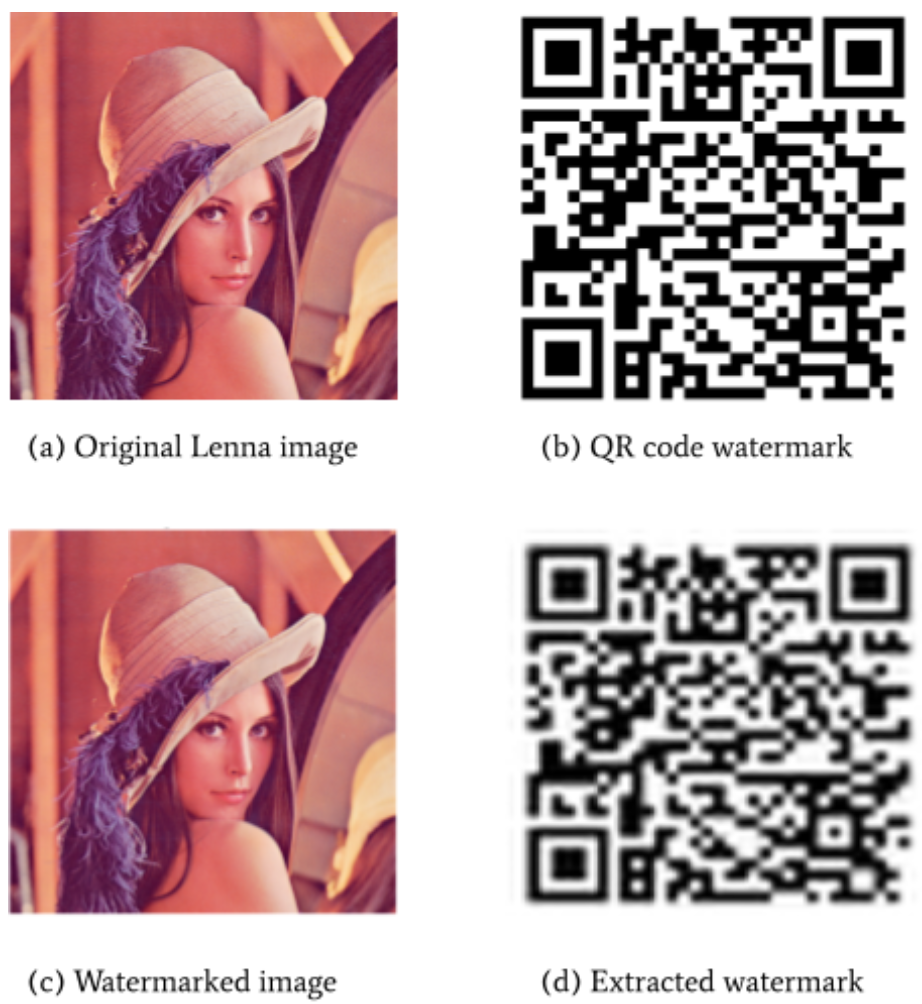}}
\caption{Watermark embedding and extracting}
\label{fig_watermark_embedding_and_extracting}
\end{figure}

In step 4, the proposed framework generates a QR code referring to the blockchain’s created block and utilizes the block hash value to generate a 64x64 QR code. The generated QR code serves as the watermark. The QR code watermark in Figure~\ref{fig_watermark_embedding_and_extracting} (b) is created referencing the block hash (i.e., \texttt{0x8065f1948301a8ab627b58346629df699912db507eb71beec0677281e55bb2d1} referring to Table~\ref{tab_blockchain}).

\subsubsection{Digital watermark embedding and extracting}

In step 5, the proposed framework utilizes the generated QR Code as the digital watermark. The Watermark Encoder embeds this digital watermark into the original digital image, creating the digitally watermarked image (content) for public sharing. The frequency-domain digital floating watermark algorithm, based on discrete cosine transform (DCT), is employed to embed the digital watermark into the original digital image content. The inverse transform (IDCT) is subsequently used to extract the watermark when asserting ownership is performed, including tamper detection. In the proposed approach, the framework is implemented using the two-dimensional DCT and two-dimensional IDCT.

For example, Figure~\ref{fig_watermark_embedding_and_extracting} (c) illustrates the outcome after the watermark encoding step into the original Lenna image displayed in Figure~\ref{fig_watermark_embedding_and_extracting} (a) with the QR code watermark (i.e., Figure~\ref{fig_watermark_embedding_and_extracting} (b)) using the DCT algorithm. Furthermore, Figure~\ref{fig_watermark_embedding_and_extracting} (d) exhibits the extracted watermark using the IDCT algorithm in the proposed framework, a step utilized during the asserting ownership route with tamper detection.

\subsubsection{Interplanetary File System (IPFS) Content Management}
In step 6, the framework uploads the watermarked image and the corresponding block information to the IPFS network. Leveraging the IPFS content management system facilitates the seamless browsing and downloading of watermarked images (content) and pertinent copyright information within the proposed framework. This decentralized approach ensures efficient content distribution and retrieval while maintaining the integrity of ownership data and associated copyright details.

\subsubsection{Verification and Tamper Detection}

In the proposed framework, the verification and tamper detection processes are executed during the image upload stage for asserting ownership (i.e., step 5). If the image is identified as original digital content with no watermark detected, it proceeds as an initial image upload, and the framework continues with the previously discussed steps. However, to assert ownership, watermark extraction is performed using the IDCT algorithm, and the copyright information is retrieved from the block in the blockchain. Different Hash (DHA) is employed to calculate hash values for both the original image (i.e., source image from IPFS) and the watermark-detected (i.e. target) image during the tamper detection phase. Subsequently, the Hamming code distance is utilized to compare these two hash values. The comparison ensures that the hash value is consistent with the hash values stored in the blockchain and the hash value generated for the target image. This step serves as the primary verification of copyright, confirming the integrity and authenticity of the digital content.

Further, copyright violation detection is processed through tamper detection. The proposed framework adopts a layered passive tamper detection approach to detect tampering against various attacks comprehensively. The  tamper detection implementation consists of four independent layers:

\paragraph{The Structural Similarity Index (SSIM) Evaluation}
The SSIM\cite{Zhou2004SSIM} serves as a perceptual metric that quantifies image quality degradation resulting from processes like data compression or losses during data transmission in the first layer. In our framework, the tampered image (i.e., uploaded image) is considered the processed image, while the original image (i.e., image downloaded from the IPFS network) serves as the reference. The SSIM is employed to measure the similarity between the two images, facilitating easy comparison.

\paragraph{The Matching Feature Result (MFR) Evaluation}
The second layer focuses on measuring matching objects between images. The framework employs the Oriented Fast and Robust Brief (ORB) algorithm to identify feature points and calculate hamming distances. The Matching Feature Result (MFR) is expressed by;
\begin{equation}
MFR = \frac{Total\; matching\; key\; points}{Total\; key\; points\; in\; tampered\; image}\times 100 \%
\label{eq_histogram_comparison}
\end{equation}


\paragraph{Histogram Comparison}
An image histogram provides a graphical representation of pixel intensity distribution in a digital image. The proposed framework incorporates histogram comparison in tamper detection in the third layer, revealing property variations within the tampered original image.

\paragraph{Marking tampered areas}
The proposed framework employs template matching, a technique in digital image processing, to identify areas in the original image that match a template image, effectively locating modified areas in the fourth layer. A Masking approach is used with the SSIM algorithm to detect the embedding of text-based content in the tampered areas.

\section{Experiments}
\label{lbl:experiments}
We evaluated the SecureRight framework from two perspectives: (a) Assessing the robustness in extracting the blind digital watermark under different attacks. (b) Evaluating the accuracy in Tamper Detection under typical attacks.
\subsection{Experimental setup}

\subsubsection{Digital Watermark Attacks}
\label{sub_sec_digital_watermark_attacks}
The robustness of any watermarking scheme is a critical factor, particularly in the face of various attacks known to affect watermarking systems. To thoroughly assess the robustness of the proposed SecureRight framework, we subjected to the following types of digital watermark attacks:
\begin{itemize}
    \item \textit{Color Attack} - Alters colour levels intentionally to create visual changes, deceiving viewers without significant structural modifications.
    \item \textit{Histogram Attack} - Estimates a watermark by analyzing only the histogram of a subjected image, manipulating or removing specific features related to the watermark.
    \item \textit{Blur Attacks} - Introduce various blurring levels, challenging watermark extraction or detection. (1) The Medium blur attack applies moderate blurring. (2) The Blur attack deliberately blurs the entire image or specific portions. (3) The Gaussian blur Attack smooth pixel intensities using a Gaussian filter.
    \item \textit{Erase Attack} - Involves intentional removal of pixels or regions, resulting in notable distortion measured by mean square error (MSE).
    \item \textit{JPEG Attack} - Undermines robustness by adjusting the "quality factor" during compression, balancing image quality reduction and maintaining visual plausibility.
    \item \textit{Salt Pepper and Gaussian Attacks} - The Salt and Pepper Attack introduces random disturbances, darkening some pixels (pepper) and brightening others (salt), mimicking noise addition. At the same time, the Gaussian attack applies Gaussian blur to the entire image or specific regions, smoothing it by averaging pixel values and reducing fine details.
\end{itemize}

\subsubsection{Tampering Techniques}
\label{sub_sec_tampering_techniques}
Tampering techniques encompass a variety of manipulative actions aimed at altering or deceiving the content of digital images. In evaluating copyright violation, the framework must demonstrate the capability to detect standard tampering techniques, showcasing its robustness in tamper detection.

\begin{itemize}
    \item \textit{Copy-Move} - Copy-move forgery involves copying and pasting a specific part of an image within the same image to conceal or replicate objects, creating the illusion of multiple occurrences
    \item \textit{Text Image-Splicing} - Text image-splicing manipulates images by adding or altering textual content, achieved by copying and pasting parts of an image onto another.
    \item \textit{Resize} - The resize attack alters image size through geometric transformations, either enlarging or shrinking the image dimensions, impacting its overall visual appearance.
    \item \textit{Cropping} - The cropping attack trims portions of an image, reducing the canvas size or eliminating specific regions along the borders to exclude unwanted details or elements.
    \item \textit{Noising and Blurring} - The noise and blurring attack introduces random pixel variations (noise) or blurring effects to degrade visual quality and alter image characteristics, making subtle changes or additions challenging to detect.
\end{itemize}

\subsubsection{Data}
To assess the performance and functionality of the proposed system, we employed a standard test image (512x512 px) widely recognized in image processing. This image, commonly known as Lenna or Lena ( i.e., Figure~\ref{fig_watermark_embedding_and_extracting} (a)), serves as a benchmark for evaluating various image processing algorithms and techniques.


\subsubsection{Evaluation Measurements}
To assess the effectiveness of the proposed framework in terms of copyright protection and tamper detection, we have employed the following evaluation metrics.

\paragraph{Mean Square Error (MSE)}
We evaluated the quality of the watermark embedding task in the proposed system by calculating the Mean Squared Error (MSE) between the original image and the watermarked image. A lower MSE indicates a higher quality of watermark generation. The MSE is computed using the formula:
\begin{equation}
    MSE = \frac{1}{mn}\displaystyle\sum\limits_{i=0}^{m-1}\displaystyle\sum\limits_{j=0}^{n-1} [I(i,j)-K(i,j)]^2\label{eq_mse}
\end{equation}
where $I$ represents the noise-free $m \times n$ monochrome image, $K$ is its noisy approximation, and $m$ and $n$ are the dimensions of the image.

\paragraph{Peak Signal to Noise Ratio (PSNR)}

To evaluate watermark transparency, our approach employs the Peak Signal-to-Noise-Ratio (PSNR) between the original and watermarked images. A higher PSNR indicates a lower likelihood of perceiving the watermark. PSNR is used for evaluating the image quality, which is calculated using the formula:
\begin{equation}
    PSNR = 20. \log_{10}{(MAX) - 10. \log_{10}{(MSE)}}\label{eq_psnr}
\end{equation}
Here, $MAX$ represents the maximum value of the original image pixels, and $MSE$ is the Mean Squared Error between the original and watermarked images.

\paragraph{Bit Error}
The quality of watermark extraction in the proposed system was assessed by quantifying the number of bit errors between the original and extracted watermark. A lower bit error count signifies a higher quality of watermark restoration.

\paragraph{Structural Similarity Index (SSIM)}

We utilize the Structural Similarity Index (SSIM) to evaluate watermarked image quality during tamper detection. The SSIM produces a value between 0 and 1, where 1 indicates perfect structural similarity between two images. The SSIM is calculated using the formula:
\begin{equation}
    SSIM(X,Y) = \frac{(2u_{X}u_{Y}+c_{1})(2\sigma_{XY}+C_{2})}{(u_{X}^2+u_{Y}^2+c_{1})(\sigma_{X}^2+\sigma_{Y}^2+c_{2})}\label{eq_ssim}
\end{equation}
Here $u_{X}$ and $u_{Y}$ are the mean values of images $X$ and $Y$, 
$\sigma X$  and $\sigma Y$ are the standard deviations of images $X$, and $Y$, $\sigma XY$ is the covariance of images $X$ and $Y$ and $C_{1}$ and $C_{2}$ are two variables used to stabilize the division with a weak denominator.

\subsection{Results}

\subsubsection{Digital Watermark Extraction}

In evaluating the robustness of extracting the blind digital watermark, we thoroughly assessed the Discrete Cosine Transform ($DCT$) algorithm used in the proposed framework against various common digital watermark attack types. Additionally, a cross-evaluation was conducted, comparing the results with two other algorithms: $DWT\_DCT$ (Discrete Wavelet Transform $+$ Discrete Cosine Transform) and $DCT\_SVD$ (Discrete Cosine Transform $+$ Singular Value Decomposition). The evaluation aimed to verify the capability of extracting the watermark from attacked images, testing across nine digital watermark attacks as discussed in section~\ref{sub_sec_digital_watermark_attacks}.

\begin{figure*}
\centerline{\includegraphics[width=0.88\textwidth]{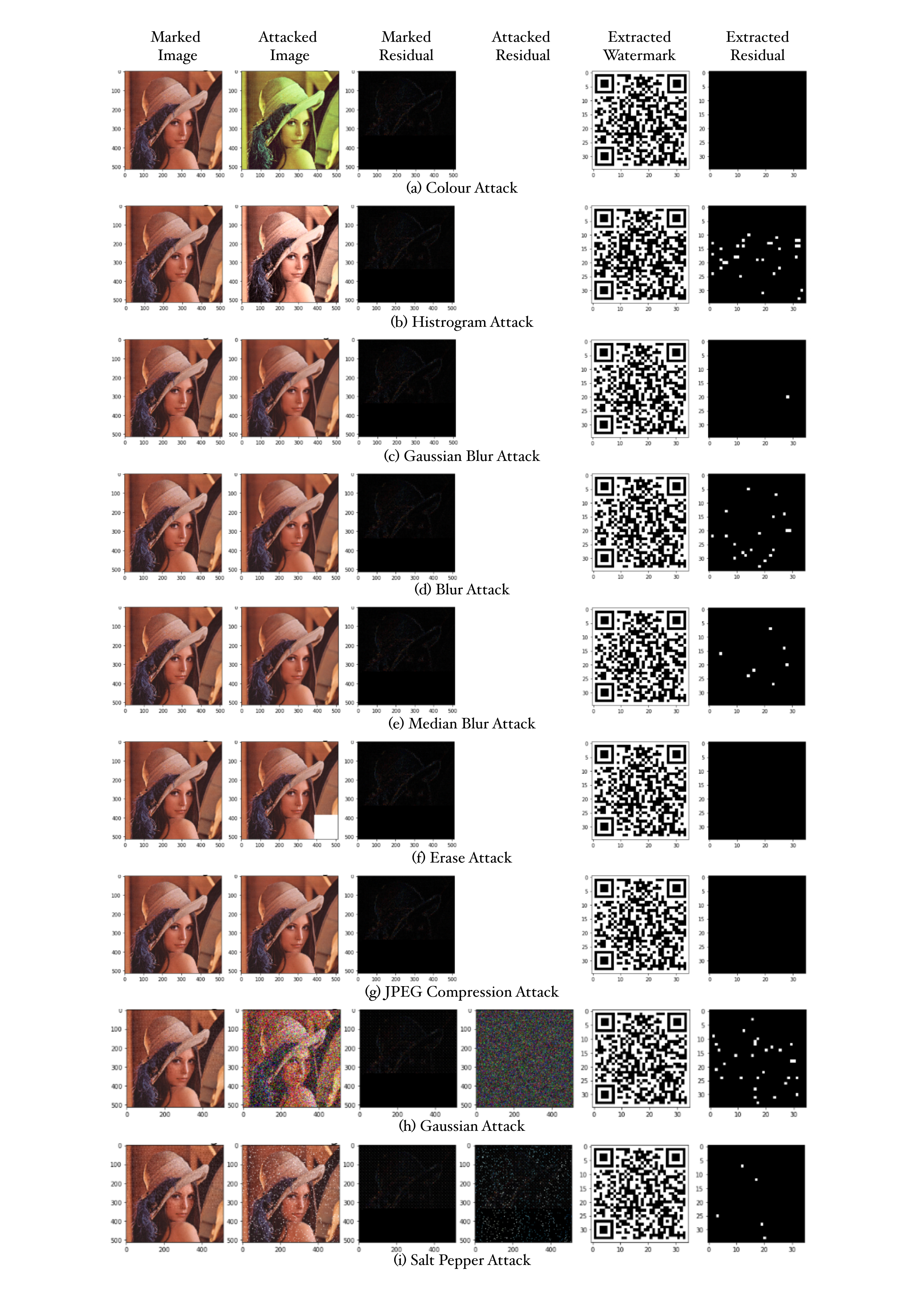}}
\caption{Experiment on digital watermark extraction against attacks using DCT algorithm}
\label{fig_watermark_attacks}
\end{figure*}

The results presented in Figure~\ref{fig_watermark_attacks} illustrate the capability of watermark recognition and successful recovery of the QR code. Table 2 summarizes the experiment results, demonstrating the validation and performance of the selected $DCT$ algorithm within the intended task of the proposed framework. The adoption of the $DCT$ algorithm for watermarking processes (embedding and extraction) consistently produced low PSNR values while maintaining a minimal number of bit errors across various attacks, as it indicates a high likelihood of perceiving the watermark compared to DWT\_DCT and DCT\_SVD algorithms.

The key findings from the results are as follows:
\begin{itemize}
    \item $DCT$ significantly reduces embedded image quality, but it greatly improves the robustness of the watermark, making it suitable for most scenarios.
    \item $DWT\_DCT$ exhibits relatively poor robustness, making it more suitable for situations where the image is minimally distorted.
    \item $DCT\_SVD$ demonstrates a significant improvement in image quality compared to other $DCT$ methods but is less robust to $DCT$ concerning digital watermark attacks.
\end{itemize}

\begin{table*}[htbp]
\caption{Experiment of peformance of DCT, DWT\_DCT and DCT\_SVD extraction algorithms with digital watermark attacks for Lenna test image}
\begin{center}
\begin{tabular}{llllll}
\hline
\textbf{Attack Type}    & \textbf{Image}                 & \textbf{Extraction Algorithm} & \textbf{MSE \textdownarrow}             & \textbf{PSNR (dB) \textdownarrow}          & \textbf{Bit errors \textdownarrow} \\ \hline
\multirow{3}{*}{Colour }    & \ref{fig_watermark_attacks}(a) & DCT             & 0.0001260515            & 43.5626606278 & 0        \\   
                                  & & DWT\_DCT        & 0.0000090726          & 49.2767217046         & 594                 \\   
                                  & & DCT\_SVD        & 0.000001857 & 52.7214147791           & 0
                                  \\ \midrule
\multirow{3}{*}{Histogram }  & \ref{fig_watermark_attacks}(b) & DCT             & 0.0001260515            & 43.5626606278 & 310        \\   
                                  & & DWT\_DCT        & 0.0000090726           & 49.2767217046         & 740                 \\   
                                  & & DCT\_SVD        & 0.000001857 & 52.7214147791           & 447                 \\ \midrule 
\multirow{3}{*}{Gaussian Blur}     & \ref{fig_watermark_attacks}(c) & DCT             & 0.0001260515            & 43.5626606278 & 449                 \\   
                                  & & DWT\_DCT        & 0.0000090726           & 49.2767217046         & 720                 \\   
                                  & & DCT\_SVD        & 0.000001857 & 52.7214147791           & 73        \\ \midrule 
\multirow{3}{*}{Blur}              & \ref{fig_watermark_attacks}(d) & DCT             & 0.0001260515            & 43.5626606278 & 451                 \\   
                                  & & DWT\_DCT        & 0.0000090726           & 49.2767217046         & 706                 \\   
                                  & & DCT\_SVD        & 0.000001857 & 52.7214147791           & 116        \\ \midrule 
\multirow{3}{*}{Median Blur}       & \ref{fig_watermark_attacks}(e) & DCT             & 0.0001260515            & 43.5626606278 & 444                 \\   
                                  & & DWT\_DCT        & 0.0000090726           & 49.2767217046         & 731                 \\   
                                  & & DCT\_SVD        & 0.000001857 & 52.7214147791           & 163        \\ \midrule
\multirow{3}{*}{Erase}      & \ref{fig_watermark_attacks}(f) & DCT             & 0.0001260515            & 43.5626606278 & 0          \\   
                                  & & DWT\_DCT        & 0.0000090726           & 49.2767217046         & 16                  \\   
                                  & & DCT\_SVD        & 0.000001857 & 52.7214147791           & 16                  \\ \midrule 
\multirow{3}{*}{JPEG Compression}   & \ref{fig_watermark_attacks}(g)    & DCT             & 0.0001260515            & 43.5626606278 & 0          \\   
                                  & & DWT\_DCT        & 0.0000090726           & 49.2767217046         & 193                 \\   
                                  & & DCT\_SVD        & 0.000001857 & 52.7214147791           & 0        \\ \midrule 
\multirow{3}{*}{Gaussian}      & \ref{fig_watermark_attacks}(h)    & DCT             & 0.0001260515            & 43.5626606278 & 0          \\   
                                  & & DWT\_DCT        & 0.0000090726           & 49.2767217046         & 459                 \\   
                                  & & DCT\_SVD        & 0.000001857 & 52.7214147791           & 2                   \\ \midrule 
\multirow{3}{*}{Salt Pepper}    & \ref{fig_watermark_attacks}(i)   & DCT             & 0.0001260515            & 43.5626606278  & 10         \\   
                                  & & DWT\_DCT        & 0.0000090726          & 49.2767217046         & 533                 \\   
                                  & & DCT\_SVD        & 0.000001857 & 52.7214147791           & 448   
          \\ \hline
\end{tabular}
\label{tab_watermark_resutls}
\end{center}
\end{table*}

\subsubsection{Tamper Detection}

To assess the quality of tamper detection, we employed the Structural Similarity Index (SSIM) (i.e., equation~\ref{eq_ssim}) as a quantifier, which reflects the similarity between the original image and the tampered image across various attack types described in section~\ref{sub_sec_tampering_techniques}. Additionally, we measured the matching objects between the target and the original images using the Matching Feature Result (MFR) (i.e., equation~\ref{eq_histogram_comparison}), considering feature points. The evaluation of MFR was extended to include a comparison of histograms.

\begin{figure*}[htbp]
\centerline{\includegraphics[width=0.9 \textwidth]{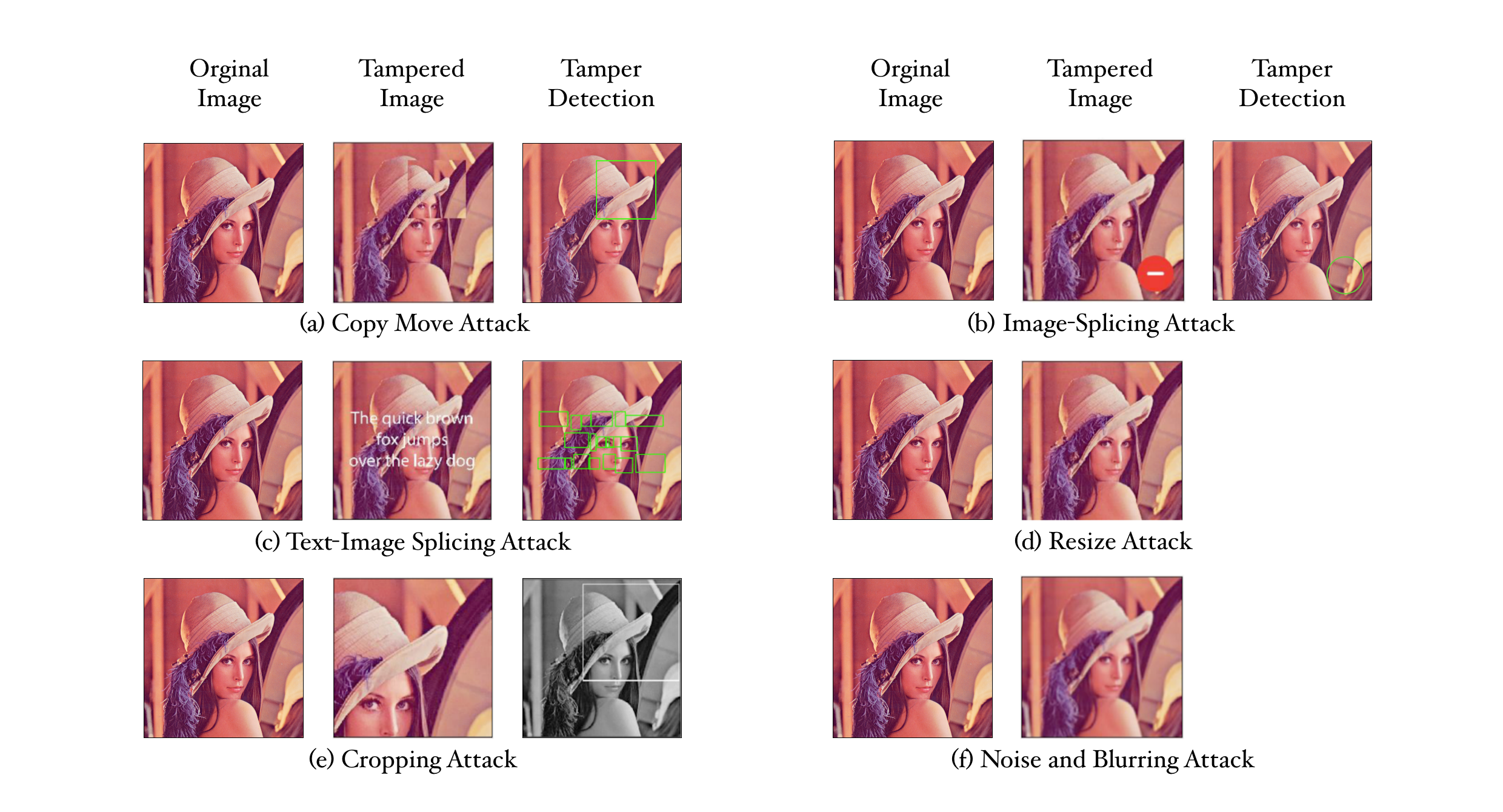}}
\caption{Experiment on tamper detection visualisation (Layer 4) against attacks}
\label{fig_tamper_detection}
\end{figure*}

Tamper detection was evaluated based on the six stated attacks and the results from Layers 4 (Visualization Layer) shown in Figure~\ref{fig_watermark_attacks}, while the results from Layers 1 to 3 are summarized in Table \ref{fig_tamper_detection}.

The proposed layered approach for tamper detection and measurement enhances tamper detection accuracy in the framework. As reflected in Table \ref{tab_tamper_resutls}, the collaboration of different layers provides better insights into tamper detection for standard types of attacks.

\begin{table}[htbp]
\caption{Layered Tamper detection results against attacks}
\begin{center}
\begin{tabular}{lllll}
\hline
\textbf{Attack}  & \textbf{Image}      & \textbf{\begin{tabular}[c]{@{}l@{}}Layer 1 SSIM \textuparrow \end{tabular}} & \textbf{\begin{tabular}[c]{@{}l@{}}Layer 2 MFR \textuparrow (\%)\end{tabular}} & \textbf{\begin{tabular}[c]{@{}l@{}} Layer 3  Histogram MFR \textuparrow (\%)\end{tabular}} \\ \hline
Copy-move & \ref{fig_tamper_detection}(a)                 & 0.9007              & 7.57              & 93.33                       \\
Image-splicing &  \ref{fig_tamper_detection}(b)              & 0.9638              & 3.57              & 99.33                        \\ 
Text Image-splicing  & \ref{fig_tamper_detection}(c)       & 0.9156              & 6.63                 & 99.65                            \\
Resize  & \ref{fig_tamper_detection}(d)                      & 0.9017              & 54.51             & 74.04                        \\
Cropping & \ref{fig_tamper_detection}(e)                    & 0.6138              & 96.51          & 47.00                         \\
Nosing and blurring & \ref{fig_tamper_detection}(f)         & 0.8431              & 83.67                 & 92.33                       \\ \hline
\end{tabular}
\label{tab_tamper_resutls}
\end{center}
\end{table}
 
The proposed framework demonstrates robust support for digital watermarking, certifying the authentication of internet image sources and detecting misuse, even in various attacks. Experimental results indicate the practicality and effectiveness of the framework, showcasing its ability to identify tampered content and trace individuals who misuse or violate confidential internet pictures, addressing potential copyright violations.

\section{Conclusion}
\label{lbl:conclusion}
This paper introduces a blockchain-based copyright management system integrating digital watermarks, blockchain technology, perceptual hashing, QR codes, and the InterPlanetary File System (IPFS). The blockchain securely stores watermark information, providing timestamp authentication for multiple watermarks and confirming their chronological order. Perceptual hashing simplifies watermark extraction by generating a hash based on the image's structural information, enhancing digital watermark robustness. The hash and copyright message are concealed in a QR code, contributing to overall watermark robustness. IPFS is used for decentralized storage and distribution of watermarked images, eliminating the need for a central server. This comprehensive approach improves the efficacy of digital watermarking technology in copyright protection, enabling copyright management and publication in peer-to-peer networks without a trusted third party. Network security is maintained through cryptographic identity confirmation among nodes, reducing risks such as information leakage and data corruption. The solution accelerates publishing copyrighted works online, fostering circulation and enhancing copyright protection for multiple creations. Further enhancements, including advanced tamper detection methods and extensions to multimedia file types, can be explored for improved results. The proposed system records and proves the copyright information of each owner during the creative process, safeguarding the legitimate rights of copyright owners. While the current focus is on digital images, the framework can be expanded to encompass audio, video, and other multimedia file types, creating diverse copyright management systems.

\bibliographystyle{ACM-Reference-Format}

\end{document}